\documentstyle[12pt,draft]{article}

\newcommand{\be}{\begin{equation}}
\newcommand{\ee}{\end{equation}}
\newcommand{\ba}{\begin{eqnarray}}
\newcommand{\ea}{\end{eqnarray}}
\newcommand{\baa}{\begin{eqnarray*}}
\newcommand{\eaa}{\end{eqnarray*}}
\newcommand{\ci}[1]{\cite{#1}}

\newcommand{\lab}[1]{\label{#1}}
\newcommand{\re}[1]{(\ref{#1})}

\newcounter{hran}

\newcommand{\ah}{\ifmmode\hat{a}\else$\hat{a}$\fi}
\newcommand{\ahd}{\ifmmode\hat{a}^\dagger\else$\hat{a}^\dagger$\fi}
\newcommand{\ad}{\ifmmode a^+\else$a^+$\fi}
\newcommand{\cd}{c^+}
\newcommand{\bx}{\ifmmode{\mbox{\bf x}}\else{\bf x}\fi}
\newcommand{\by}{\ifmmode{\mbox{\bf y}}\else{\bf y}\fi}
\newcommand{\dl}{\ifmmode\delta\else$\delta$\fi}
\newcommand{\Dl}{\ifmmode\Delta\else$\Delta$\fi}
\newcommand{\al}{\ifmmode\alpha\else$\alpha$\fi}
\newcommand{\ph}{\ifmmode\hat\varphi\else$\hat\varphi$\fi}
\newcommand{\vp}{\ifmmode\varphi\else$\varphi$\fi}
\newcommand{\pl}{\ifmmode\partial\else$\partial$\fi}
\newcommand{\kp}{\ifmmode\kappa\else$\kappa$\fi}
\newcommand{\tet}{\ifmmode\theta\else$\theta$\fi}
\newcommand{\eps}{\ifmmode\varepsilon\else$\varepsilon$\fi}
\newcommand{\derl}[1]{\ifmmode\frac{\stackrel{\rightarrow}{\pl}}{\pl #1}
        \else$\frac{\stackrel{\rightarrow}{\pl}}{\pl #1}$\fi}
\newcommand{\derr}[1]{\ifmmode\frac{\stackrel{\leftarrow}{\pl}}{\pl #1}
        \else$\frac{\stackrel{\leftarrow}{\pl}}{\pl #1}$\fi}
\newcommand{\der}[1]{\ifmmode\frac{\pl}{\pl #1}
	\else$\frac{\pl}{\pl #1}$\fi}
\newcommand{\dertwo}[2]{\ifmmode\frac{\pl #1}{\pl #2}
	\else$\frac{\pl #1}{\pl #2}$\fi}
\newcommand{\N}{\ifmmode{\mathrm N}\else {\mathrm N}\fi}
\newcommand{\T}{\ifmmode{\mathrm T}\else {\mathrm T}\fi}
\newcommand{\Sym}{\ifmmode{\mathrm Sym}\else {\mathrm Sym}\fi}
\newcommand{\PPr}{\ifmmode{\stackrel{\leftarrow}{\cal P}}
	\else $\stackrel{\leftarrow}{\cal P}$\fi}

\newcommand{\tr}{\mbox{tr}\,}

\newcommand{\dd}{{\mathrm d}}

\newcommand{\laa}{\langle\!\langle}
\newcommand{\raa}{\rangle\!\rangle}

\newcommand{\E}{{\mathrm e}}

\title{$q$-Functional Field Theory for particles with exotic statistics}
\author{
K.N.Ilinski $^{1,2,3}$
\thanks{E-mail: kni@th.ph.bham.ac.uk}, \
G.V.Kalinin $^{1}$
\thanks{E-mail: kalinin@snoopy.phys.spbu.ru} \
and A.S.Stepanenko $^{1,2,4}$
\thanks{E-mail: ass@th.ph.bham.ac.uk}\\ [0.2cm]
{\small\it $^{1}$ IPhys Group, CAPE, St.-Petersburg State
University,}\\
{\small\it 14-th line of V.I. 29, St-Petersburg, 199178, Russian
Federation}
\\ [0.2cm]
{\small\it $^{2}$ School of Physics and Space Research,
University of Birmingham,} \\
{\small\it Edgbaston B15 2TT, Birmingham, United Kingdom}
\\ [0.2cm]
{\small\it $^{3}$ Theoretical Department, Institute of Spectroscopy,
Troitsk,\ \ \ \ \ \ \ \ \ \ \ } \\
{\small\it  \ \ \   Moscow region, 142092, Russian Federation }
\\ [0.2cm]
{\small\it $^{4}$ Theoretical Department, St.-Petersburg Nuclear Physics
Institute,}\\
{\small\it  Gatchina, St.-Petersburg, 188350, Russian Federation}
}
\date{  }

\begin{document}
\maketitle
\vskip -9.5cm
\vskip 9.5cm

\begin{abstract}
In the paper we give consecutive description of functional methods of
quantum field theory for systems of interacting $q$-particles. These
particles obey exotic statistics and appear in many problems of condensed
matter physics, magnetism and quantum optics. Motivated by the general
ideas of standard field theory we derive formulae in $q$-functional
derivatives for the partition function and Green's functions generating
functional for systems of exotic particles. This leads to a corresponding
perturbation series and a diagram technique. Results are illustrated by a
consideration of an one-dimensional $q$-particle system and compared with
some exact expressions obtained earlier.
\end{abstract}

\section{Introduction}
In this paper we deal with so-called $q$-particles, {\it i.e.} particles
which appear as a result of quantization of a Hamiltonian classical
dynamics on $q$-deformed graded-commutative algebras~\cite{BI}. Their
creation and annihilation operators $\ahd_{k} ,\ah_{k}$ obey the
following commutation relations:
\ba
&&{}\ah_{k}\ah_{j} =
\kappa q_{kj}\ah_{j}\ah_{k}\ ,\quad
\ahd_{k}\ahd_{j} =
\kappa q_{kj}\ahd_{j}\ahd_{k}\ ,\quad
\ah_{k}\ahd_{j} =
\kappa q^{-1}_{kj}\ahd_{j}\ah_{k} +\delta_{kj}\ ,
\nonumber\\&&
\lab{com-rel}
\ah_{k}\ah_{k} =
\ahd_{k}\ahd_{k} = 0 \quad
\mbox{for $q$-fermions}\ .
\ea
Here the deformation parameters $q_{kj}$ possess the property
$$
q_{kj} = \E^{i\phi_{kj}}\ ,\qquad
\phi_{kj}=-\phi_{jk}\ ,\quad \phi_{kj}\in{I\!\!R}
$$
and $\kp$ serves to unify formulae for deformed bosonic and
deformed fermionic cases. As usual it has the form:
$\kp=+1$ for $q$-bosons and $\kp=-1$ for $q$-fermions.

The notion of the exotic quantum statistics seems to be rather artificial
or mathematical. Number of papers where particles with the statistics
emerged in a context of the parastatistics~\cite{BI,Mar,F}, $q$-extended
Supersymmetry~\cite{IU1,qIS}, Parasupersymmetry~\cite{IU1,BD} and other
similar problems apparently prove the statement. However, solid state
physics, quantum optics and theory of magnetics give examples of other
kind.  Indeed, anyons (particles with exotic braiding statistics) are
important in some attempts to understand the physical features of planar
systems~\cite{Fr}, the quantum Hall effect~\cite{QH} and high temperature
superconductivity~\cite{sc}.  In contrast to these examples where anyons
serve as auxiliary objects for the construction of one of possible
scenarios  there is a wide field in the quantum nonlinear optics where
$q$-particles are the main components.  This is a theory of the
collective behavior of excitons with small radius (Frenkel Excitons and
Charge-Transfer Excitons (CTE)) \cite{A1}. The studies investigate
possibilities of formation of the Frenkel biexcitons and the observation
of phase transitions in exciton systems in molecular crystals (Bose-
Einstein condensation of excitons \cite{AT}, structural phase transitions
in crystals with high excitonic concentrations, dielectric-metal phase
transition in a system of interacting CTE \cite{AI} and others). Strictly
speaking excitons are not particles. They are quasiparticles describing
molecular excitations and are of great  importance in the analysis of
nonlinear optical processes which accompany propagation of high-intensity
light fluxes whose frequencies are in the range of the exciton absorption
bands \cite{A2}.  These excitons obey exotic statistics (Pauli
statistics) \cite{A0} coinciding with $q$-particles statistics for
$q=-1$.  The general case of $q=e^{i\phi}$ arises if we try to take
into account phenomenologically some nonlinear effects (such as the
difference in the creation time of molecular excitations for different
types of molecules).  This effect can be modeled by the change of the
Paulion commutation relations to those of the $q$-particles using the
method developed in \cite{PS}. Noteworthy, even the investigation of the
behavior of low dimensional exciton systems is meaningful. The best
example is the exact solution for one-dimensional Paulion chain
\cite{LMS} caused great advances in the theory of the so called
J-aggregates, {\it i.e.} molecular aggregates with unusually sharp
absorption band (\cite{KS} and Refs.  therein). The investigations of
exciton systems on interfaces closely connect with the successes of
contemporary technology.  All these show that $q$-particles find deep
applications in modern physical theories and motivate our objective to
derive the appropriate field theoretical technique for them.

Recently, it was shown that $q$-functional form of Wick's theorems for
creation and annihilation operators of the $q$-particles can be
formulated and {\it they have the same formal expressions as fermionic
and bosonic ones but differ by a nature of fields}\cite{IKS1}. It means
that in the case of the $q$-particles certain $q$-deformed algebras
should be used exactly as it was with Grassmann algebra in the case of
fermions or the complex numbers in the case of bosons. This fact allows
us in the present paper construct a machinery of the quantum field theory
for the exotic particles going along the way of standard textbooks.

In a sense, the present work may be considered as a consequential step to
the quantum field theory for the exotic particles which follows from
previous papers \cite{BI,IS,IKS1}. Indeed, a construction of classical
and quantum dynamics on graded-commutative spaces~\cite{BI} stated a
connection of the $q$-particles and the $q$-deformed classical variables.
These variables then have been used to introduce the corresponding
coherent states and to derive $q$-functional integrals~\cite{IS} for
$q$-particle systems. It is well-known that Wick's theorems present other
way to the functional integrals and field-theoretical methods. Since the
$q$-functional Wick's theorems have been proved~\cite{IKS1} we now make
the ends met and derive the field-theoretical technique and functional
integrals from the theorems.  This step (this paper) completes the
program.

The paper is organized as follows. In the next section we give a
functional representation for the partition function and Green's
functions generating functional. Then in the sections 3 we illustrate the
developed technique with an example of calculations for simple
a one-dimensional $q$-fermionic system. Earlier, in Ref.\cite{IK} we
presented exact results for the partition function and Green's functions
for the system using a functional integral method. This gives us a
possibility to compare these two approaches and to state their
consistency. Closing remarks conclude the paper.

\section{$q$-Functional field theory}

In the quantum statistics all equilibrium physical properties are
described by the density matrix given by the operator
\be
\rho=\exp[-\beta H]\ ,\qquad \beta=1/kT\ .
\lab{1}
\ee
In particular,
the equilibrium thermodynamics is governed by the partition function
which is a trace of the density matrix:
\be
Z=\tr \rho\ .
\lab{2}
\ee
Mean value $(\!(\hat b)\!)$ for
an arbitrary quantum operator $\hat b$ then
may be calculated as
\be
(\!(\hat b)\!)=Z^{-1}\tr [\rho\hat b]\ .
\lab{3}
\ee
In general, the partition function cannot be calculated exactly and
one should develop a perturbation theory. To do this, as usual we need to
divide the Hamiltonian into two parts: so-called ``free'' Hamiltonian
$H_0$ which may be treated exactly and an interaction $V$ considered as a
perturbation:
\be
H=H_0+V\ ,\qquad H_0 = \sum_{k} \eps_k \ahd_k\ah_k\ .
\lab{H0}
\ee
Since there is no appropriate way to treat hopping terms exactly
we include them in the interaction.

The ratio of $\rho$-matrices $\rho^{-1}_0\rho=\exp[\beta H_0]\exp[-\beta
H]$ coincides with the Euclidean evolution operator $U(\beta,0)$ in the
interaction representation. Hence, the partition function and various
averages require explicit expression for the operator $U(\beta,0)$ in
some approximation and they may be written as
\be
Z=\tr\rho=\tr[\rho_0 U(\beta,0)]=Z_0\laa U(\beta,0)\raa\ ,
\lab{Z-Z0}
\ee
where
\be
\laa \hat b\raa=Z_0^{-1}\tr[\rho_0 \hat b]\ ,\qquad
Z_0=\tr[\rho_0]\ .
\lab{18}
\ee
Creation and annihilation operators $\ahd, \ah$ in
the interaction representation take the form
\be
\ahd_k(\tau)= \E^{\tau H_0} \ahd_k\E^{-\tau H_0}\ ,\qquad
\ah_k(\tau) = \E^{\tau H_0} \ah_k \E^{-\tau H_0}\,.
\lab{t}
\ee
From (\ref{H0}) we get an explicit form for the evolution of the operators:
$$
\ahd_k(\tau) = \ahd_k \E^{\tau\eps_k}\ ,\qquad
\ah_k(\tau) = \ah_k \E^{-\tau\eps_k}\,.
$$
So the problem is reduced to the calculation of the evolution operator
$U(\tau_1,\tau_2)$ in the interaction representation. It is defined by the
following equation:
\be
\frac{\pl U(\tau_1,\tau_2)}{\pl\tau_1}  = -V(\tau_1)
U(\tau_1,\tau_2)\ ,
\lab{5}
\ee
with the Hamiltonian in the interaction representation
$$
V(\tau) = \E^{H_0\tau}V\E^{-H_0\tau} \ .
$$
The solution of eq.~\re{5} with the initial condition $U(\tau,\tau)={\rm
I}$ is the Volterra series:
\ba
U(\tau_1,\tau_2) &=&\E^{H_0\tau_1}
\E^{H(\tau_2-\tau_1)}\E^{-H_0\tau_2}=
\nonumber\\
&=&\sum^\infty_{n=0}(-1)^n\int_{\tau_2}^{\tau_1}\!\dd t_1\dots
\int_{\tau_2}^{\tau_1}\!\dd t_n\
\theta(1\dots n)V(t_1)\dots V(t_n)\ .
\lab{series-Volt}
\ea
If the operator $V$ is an operator functional of the bosonic type then
RHS of the equation can be represented as the standard Dyson T$_{\rm
D}$-exponent by the symmetrization with respect to the permutation of
time variables $t_k$.  In contrast to the undeformed case this condition
holds only for very special cases. In general, the symmetrization of
RHS give us another type of T-exponent. But this type of T-exponent is
not consistent with nature of operators which makes difficult to deal
with it. It would be natural to try to $q$-symmetrize RHS of
(\ref{series-Volt}) but different monomials in the operator $V$ are
permuted in different ways. So we can not $q$-symmetrize the Volterra
series with respect to the permutation of the whole operator functionals
$V$ (or time variables) and we do not obtain T$_q$-exponent. But we can
deal directly with Volterra series (\ref{series-Volt}) because each term
in the sum can be represented as T$_q$-product (due to a presence of the
$\theta$-function)~\cite{IKS1}.

Let us now reduce the Euclidean evolution operator $U(\tau_1,\tau_2)$ to
the normal form. We suppose that the operator $V$ is $q$-symmetrical one,
{\it i.e.} $V=$Sym$_qV$~\cite{IKS1} and it does not contain any time
derivatives.  Adding the sign of the $q$-chronological product to RHS of
eq.(\ref{series-Volt}) and applying Theorem~3 of Ref~\ci{IKS1} we
obtain the following rule of reducing of eq.(\ref{series-Volt}) to the
normal form:
\be U(\ahd,\ah;\tau_1,\tau_2)=\N\left[
\exp\left[\derr{a}\Dl\derr{\ad}\right]
U(\ad,a;\tau_1,\tau_2)
\Biggr|_{\stackrel{\scriptstyle \ad=\ahd}{a=\ah}}
\right]\ .
\lab{evol-norm}
\ee
where following \cite{IKS1} we have introduced classical variables
$a,\ad$ corresponding to the operators \ah, \ahd\ and satisfying the
following permutation relations
\ba
&&a_{k}a_{j} =
\kappa q_{kj}a_{j}a_{k}\ ,\quad
\ad_{k}\ad_{j} =
\kappa q_{kj}\ad_{j}\ad_{k}\ ,\quad
a_{k}\ad_{j} =
\kappa q^{-1}_{kj}\ad_{j}a_{k}\ ,
\nonumber\\
&&a_{k}a_{k} = \ad_{k}\ad_k = 0 \qquad
\mbox{for $q$-fermions}\ .
\lab{per-rel}
\ea
In formula (\ref{evol-norm}) the
$q$-chronological contraction $\Dl$ is defined by the relation
\be
\Dl=\delta_{k_1,k_2}\theta(t_1-t_2)\E^{\eps_k(t_2-t_1)} \ .
\lab{q-chron}
\ee
In the form $\derr{a}\Dl\derr{\ad}$ the summation over discrete variables
and the integration over continuous ones are implied.

Now we encounter the problem of calculating of the mean value of
an operator standing in the normal form
\be
\laa\N F(\ahd,\ah)\raa=
\mathop{F}^{\leftarrow}(\derr{c},\kappa\derr{\cd})
\laa\N\exp[\ahd c+\cd\ah]\raa\ .
\lab{34}
\ee
We imply the auxiliary fields (sources) $\cd,c$ obey permutation relations
(\ref{per-rel}).

So the problem is in calculating of the following object:
$$
f(\cd,c)=\laa\N\exp[\ahd c+\cd\ah]\raa=
\laa\exp(\ahd c)\exp(\cd\ah)\raa
$$
It is obvious that the problem falls into a set of one-dimensional ones:
\baa
f_k(\ad_k,a_k)&=&
Z^{-1}_k\tr[\exp(-\beta\eps_k\ahd_k\ah_k)
\exp(\ahd_kc_k)\exp(\cd_k\ah_k)]\\
Z_k&=&(1-\kappa\exp(-\beta\eps_k))^{-\kappa}
\eaa
The calculation reveals the results of the undeformed case and we get the
following expressions:
\be
f_k(\cd_k,c_k) = \exp(\kappa\bar n_k \ad_ka_k)\ ,\qquad
\bar n_k = \frac{\exp(-\beta\eps_k)}{1-\kappa\exp(-\beta\eps_k)}
\lab{bar-n}
\ee
Then for any operator functional $F$ it is possible to write the
following relation:
\be
\laa\N F(\ahd,\ah)\raa=
\mathop{F}^{\leftarrow}(\derr{c},\kappa\derr{\cd})
\exp[\cd d\,c]\Biggl|_{\cd=c=0}
\lab{41}
\ee
where $\displaystyle\mathop{F}^{\leftarrow}(a)$ means that in each
monomial of $F$ multipliers stand in the inverse order and the following
notation is introduced:
\be
d=\kappa\delta_{k_1,k_2}\bar{n}_k\E^{\eps_k(t_2-t_1)}
\lab{d}
\ee
Using the identity
\be
\exp[\cd d\,c]
\exp[\xi^+c + \cd\xi]=
\exp[\derr{\xi} d\derr{\xi^+}]
\exp[\xi^+ c+\cd \xi]
\lab{42}
\ee
we can rewrite \re{41} as follows
\ba
\laa\N F(\ahd ,\ah)\raa
&=&
\mathop{F}^{\leftarrow}(\derr{c},\kappa\derr{\cd})
\exp[\derr{\xi} d\derr{\xi^+}]
\exp[\xi^+ c+\cd \xi]
\Biggr|_{\stackrel{\scriptstyle\cd=c=0}{\xi^+=\xi=0}}=
\nonumber\\
&=&
\exp[\derr{\xi} d\derr{\xi^+}]
\mathop{F}^{\leftarrow}(\derr{c},\kappa\derr{\cd})
\exp[\xi^+ c+\cd \xi]
\Biggr|_{\stackrel{\scriptstyle\cd=c=0}{\xi^+=\xi=0}}=
\nonumber\\
&=&
\exp[\derr{\xi} d\derr{\xi^+}]
F(\xi^+,\xi)\Biggr|_{\xi^+=\xi=0}
\lab{mean}
\ea
The second equality is due to the fact that $d$ contains $\delta$-symbol
(it is permuted without any phase factor).
Let us introduce S-matrix functional as
\be
R(\ad,a)
 =
\exp\left[\derr{a}(d+\Dl)\derr{\ad}\right] U(\ad,a;\beta,0)\ .
\lab{s-fun}
\ee
At collecting together eqs.(\ref{Z-Z0}, \ref{evol-norm}, \ref{mean}) we
get final expression for the partition function:
\be
Z/Z_0=R(0)\ .
\lab{Z-R}
\ee
In the last formula the deformation parameter $q$ appears only in
permutation relation for variables~(\ref{per-rel}) and derivatives.

Now we consider an application of the Wick's theorems to the
calculation of Green's functions generating function.
S-matrix Green's functions (without vacuum loops)
are defined by the following relation
\be
H_n(x_1,\dots,x_n)=(\!(\T_D[\ph_{\rm
H}(x_1),\dots,\ph_{\rm H}(x_n)] )\!)
\ee
where $x\equiv(k,s,t)$, $\ph$
means $\ahd$ or $\ah$ ($\ph(k,1,t)=\ahd_k(t),\,\ph(k,2,t)=\ah_k(t)$) and
the subscript ``$H$'' means that the operator is in the Euclidean
Heisenberg representation
$$
\ph_{\rm H}(x)=\ph_{\rm H}(k,s,t)=\E^{tH}
\ph(k,s,0) \E^{-tH} \ .
$$
Using the group property of the evolution operator we obtain
$$
\ph_{\rm H}(x_1)\cdot\dots\cdot\ph_{\rm H}(x_n)=
U^{-1}(\beta,0)\,U(\beta,t_1)\ph(x_1)U(t_1,t_2)\dots\ph(x_n)U(t_n,0) \ .
$$
It means that we can write down the following equality:
\be
(\!(\T_D[\ph_{\rm H}(x_1),\dots,\ph_{\rm H}(x_n)])\!)=
\frac{\laa
\T_D[U(\beta,t_1)\ph(x_1)U(t_1,t_2)\dots\ph(x_n)U(t_n,0)]
\raa}{\laa U(\beta,0)\raa} \ .
\lab{green-inter}
\ee
Now we shall reduce the expression in the angle brackets to the normal
form and then apply eq.(\ref{mean}):
\ba
&& G_n(x_1,\dots,x_n)\equiv
\laa
\T_D[U(\beta,t_1)\ph(x_1)U(t_1,t_2)\dots\ph(x_n)U(t_n,0)]
\raa =
\nonumber\\
&&=\exp\left[\frac{1}{2}\derr{\vp}g\derr{\vp}\right]
U(\vp;\beta,t_1)\vp(x_1)U(\vp;t_1,t_2)
\dots\vp(x_n)U(\vp;t_n,0)
\Biggr|_{\vp=0} \ .
\lab{green-norm}
\ea
where matrix temperature propagator has the form
$$
g =
\left(
\begin{array}{cc}
0 &  \kappa (d + \Dl) \\
d + \Dl & 0
\end{array}
\right)
$$
These functions $G_n(x_1,\dots,x_n)$ might be obtained from
the generating functional for the Green's functions:
\be
G(A)\equiv
\left.
\exp\left[\frac{1}{2}\derr{\vp}g\derr{\vp}\right]
U(\vp,A;\beta,0)
\right|_{\vp=0}
\lab{gen}
\ee
where
\be
U(\vp,A;\beta,0) = U(\vp;\beta,0) \exp \vp A \ .
\lab{UA}
\ee
Indeed,
the Green's functions $G_n(x_1,\dots,x_n)$ are determined by the relation:
\be
G_n(x_1,\dots,x_n) =
\frac{{\displaystyle\mathop{\delta}^\leftarrow}_{\mbox{\tiny
chr}}}{\delta A(x_1)}\dots
\frac{{\displaystyle\mathop{\delta}^\leftarrow}_{\mbox{\tiny
chr}}}{\delta A(x_n)}G(A)
\lab{greenf}
\ee
where the subscript ``chr'' means a specific procedure of
the differentiation: each monomial term should be chronologically
ordered, the variable to be differentiated should be moved to the most
left position relative to other variables with the same time and then
canceled.

The formulae in the above paragraph are sufficient to work out
a field theoretical technique to calculate Green's functions and use
various tricks of a field-theoretical machinery.
However, we stop at this stage for a moment to establish a
useful connections between the generating functional (\ref{gen}) and
the S-matrix functional (\ref{s-fun}). All we need to do this is
the following easy looking formulae:
\be
\exp(-\vp A)F\left(\derl{\vp}\right)\exp(\vp A)
 = F\left(A+\derl{\vp}\right)\ ,
\lab{f1}
\ee
\be
\exp\left(A\derl{\vp}\right)F(\vp) = F(A+\vp)\ .
\lab{f2}
\ee
In these formulae we have right derivatives instead of left ones but it
is not a problem to change all formulae in a proper way.
Indeed, due to the locality of $\Delta$ and $d$ we get:
\be
R(\vp) =
\exp\left[\frac{\kappa}{2}\derl{\vp}g\derl{\vp}\right]
U(\vp;\beta,0)
\lab{s-fun1}
\ee
\be
G(A) =
\left.
\exp\left[\frac{\kappa}{2}\derl{\vp}g\derl{\vp}\right]
U(\vp,A;\beta,0)
\right|_{\vp=0}
\lab{gen1}
\ee
For the ``free" theory ($V=0$) using (\ref{f1}) we obtain
\be
G^{(0)}(A) =
\left.
\exp\left[\frac{\kappa}{2}\derl{\vp}g\derl{\vp}\right]
\exp \vp A \right|_{\vp=0}
 = \exp\left[\frac{\kappa}{2}AgA\right]   \ .
\lab{g-free}
\ee
By the same trick it is possible to rewrite the expression for
$U(\vp,A;\beta,0)$ as
\be
U(\vp,A;\beta,0) =
U\left(\kappa\derl{A};\beta,0\right) \exp\vp A \ .
\lab{UA1}
\ee
This gives us the possibility to present the
generating functional in the form:
\be
G(A) =
U\left(\kappa\derl{A};\beta,0\right)
\exp\left[\frac{\kappa}{2}AgA\right]  \ .
\lab{gen2}
\ee
From (\ref{gen1}) and (\ref{f1}) we obtain
\be
G(A) =
\left.
\exp\left[\frac{\kappa}{2}
\left(A+\derl{\vp}\right)g\left(A+\derl{\vp}\right)\right]
U(\vp;\beta,0)
\right|_{\vp=0} \ .
\lab{gen3}
\ee
As it follows from $q$-symmetry property of $\Delta,d$ and definition
(\ref{s-fun1}) the generating functional has a very simple connection
with the S-matrix functional:
$$
G(A) = \left.
\exp\left(\frac{\kappa}{2}AgA\right)
\exp\left[(gA)\derl{\vp}\right] R(\vp) \right|_{\vp=0} \ .
$$
or, using (\ref{f2}), in more compact form:
\be
G(A)
 = \exp\left(\frac{\kappa}{2}AgA\right)R(gA)
 = G^{(0)}(A) R(gA) \ .
\lab{gen4}
\ee
The inverse relation is also useful and looks as
($A\rightarrow g^{-1}A$)
\be
R(A)
 = \exp\left(-\frac{1}{2}Ag^{-1}A\right)G(g^{-1}A)\ .
\lab{s-fun2}
\ee

Let us emphasize now that {\it formulae (\ref{s-fun1}),
(\ref{gen1}), (\ref{gen4}) and (\ref{s-fun2}) are absolutely identical
to the corresponding formulae of the standard theories with the only
difference in the nature of the fields}. This is the central point to
derive a proper diagram technique.

However, before turning the attention
to the diagram rules we spend a minute to show a connection with the
$q$-functional integral formalism developed in Ref.\cite{IS}. The basis
for the bridge is eq.(\ref{gen2}).
To this end we note that the Gaussian exponent
in RHS of (\ref{gen2}) has to be expressed in a $q$-functional
integral form and then action of the differential operator
$U\left(\kappa\derl{A};\beta,0\right)$ is processed explicitly  under the
sign of the functional integral by the usual way. This gives the
complete action in the exponent under
the $q$-functional integral and, as a result, the
$q$-functional integral representation for the Green's functional
generating functional emerges.  It is exactly the same expressions which
was obtained in paper \cite{IS} for a situation of an additional
internal (anyonic) gauge field.

Let us draw outline of a diagram technique. From (\ref{gen4}) and
(\ref{s-fun2}) it obvious that a knowledge of the S-matrix functional
$R(\vp)$ is equivalent to a knowledge of all Green's functions $G_n$
and vise versa.  So we consider here the diagram technique for
the S-matrix functional only.

From the definition (\ref{s-fun}) we have the following perturbation
theory series for S-matrix:
\be
R(\vp)
 =
\exp\left[\frac12\derr{\vp}g\derr{\vp}\right]
\sum^\infty_{n=0}(-1)^n\int_{0}^{\beta}\!\dd t_1\dots
\int_{0}^{\beta}\!\dd t_n\
\theta(1\dots n)V(\vp(t_1))\dots V(\vp(t_n))\ ,
\lab{s-fun3}
\ee
It is convenient to calculate the S-matrix functional (\ref{s-fun3}) in
terms of diagrams. Each multiplier $V(\vp(t_k))$ is represented by a
vertex on line (all the vertices are ordered in time). Action of
$\derr{\vp}g\derr{\vp}$ corresponds to adding of a line $g$ connecting
a pair of vertices. The line is added by all possible ways as each
derivative $\derr{\vp}$ acts on any multiplier $V(\vp)$. In particular,
two derivatives of the quadratic form may act on the same vertex $V$.
Such lines are called tadpoles.

The result of action of the differential operation on a $n$-th term in
the sum in (\ref{s-fun3}) can be represented as a sum of diagrams
consisting of $N$ time ordered vertices with any number of added lines.
The vertex with $n$ attached lines is associated with the following
expression
\be
V_n(x_1,\dots,x_n;\vp) \equiv
\frac{\delta^n V(\vp)}{\delta\vp(x_1)\dots\delta\vp(x_n)}\ .
\lab{vn}
\ee
The arguments $x$ of the multipliers (\ref{vn}) are contracted with the
corresponding arguments of lines $g$. Multiplier $V(\vp)\equiv V_0(\vp)$
is called generating vertex. If the interaction is polynomial in fields
then finite number of multipliers (\ref{vn}) are non-zero only.

For a generic term in the sum (\ref{s-fun3}) the following representation
is valid~\cite{IKS1}:
\be
\left.
\exp\left[\frac12\sum_{ik}\derr{\vp_i}g\derr{\vp_k}\right]
(-1)^n\int_{0}^{\beta}\!\dd t_1\dots
\int_{0}^{\beta}\!\dd t_n\
\theta(1\dots n)V(\vp_1(t_1))\dots V(\vp_n(t_n))
\right|_{\vp_1=\dots=\vp_n=\vp}\ .
\lab{v1}
\ee
The diagonal terms of the quadratic form in the exponent correspond to
adding of the tadpoles. They can be accounted by introducing of a reduced
vertex
$$
V_{\mbox{\tiny red}}(\vp) =
\exp\left[\frac12\derr{\vp}g\derr{\vp}\right] V(\vp(t))\ ,
$$
and, hence, (\ref{v1}) can be rewritten as
\be
\left.
\exp\left[\frac12\sum_{i\neq k}\derr{\vp_i}g\derr{\vp_k}\right]
(-1)^n\int_{0}^{\beta}\!\dd t_1\dots
\int_{0}^{\beta}\!\dd t_n\
\theta(1\dots n)V_{\mbox{\tiny red}}(\vp_1(t_1))\dots
V_{\mbox{\tiny red}}(\vp_n(t_n))
\right|_{\vp_1=\dots=\vp_n=\vp}
\lab{v2}
\ee
The remaining terms in the differential operation add lines between
different vertices. $V(\vp)$ represents Sym$_q$-form of the interaction,
$V_{\mbox{\tiny red}}(\vp)$ does its N-form~\cite{IKS1}.

Expression (\ref{v2}) is a generic term of the perturbation theory series
and can be represented in graphic terms (diagrams). Due to the time
ordering of vertices diagram rules and, in particular, procedure of
calculation of symmetrical coefficients, differ from the standard ones
but, however, straightforward now and can be easily adapted for symbolic
computer calculations.

\section{Illustration: one-dimensional $q$-fermion system}

In the previous section we develop a general technique for a calculation
the partition function and the Green's functions generating functional.
This section is devoted to an application of the technique.  As a field
of the application we choose so-called cyclic $q$-XX-chain which was
introduced and exactly solved in Ref.\cite{IK}.  There it was shown that
the partition function and the two-point correlation functions for the
model can be calculated in explicit form so now we can use them to test
our results.  Let us remind the Hamiltonian of the $q$-XX-chain:
$$
H_0=B\sum_{m=1}^M \ahd_m \ah_m\ ,
$$
$$
V(\ahd,\ah)=A\sum_{m=1}^{M-1} (\ahd_m \ah_{m+1} + \ahd_{m+1} \ah_m) +
A(\ahd_M \ah_1 + \ahd_1 \ah_M)\ ,
$$
where the creation and annihilation operators $\ahd_m,\ah_m\,(m=1,2\dots
M)$ obey the commutation relations
\be
\begin{array}{l}
{}\ah_k \ah_j + q \ah_j \ah_{k} = 0 \ ,
\quad \ah_k \ahd_j + q^{-1} \ahd_j \ah_k = 0 \ ,\quad\qquad
q = e^{i2\pi l/N } \ ,\quad  1\leq k<j\leq n \ ,\\
{}\ah_{k} \ahd_{k} + \ahd_{k} \ah_{k}
= 1 \ ,\qquad [\ahd_k]^2 = [\ah_k]^2 = 0 \ .
\end{array}
\lab{q-p}
\ee
The relations coincide with (\ref{com-rel})
in the case $\phi_{jk}= 2\pi l/N$
for $1\le{j}<k\le{M}$.
The explicit result for the partition function of the
model has the following simple and compact form:
\begin{equation}
Z=\frac1N\sum_{n=0}^{N-1} \sum_{j=0}^{N-1} q^{-jn} \prod_{k=0}^{M-1}
\left(1+q^j \exp\biggl\{
{-\beta\Bigl[B+2A\cos(\frac{2\pi}{M}(k+\frac{(1-n)l}{N}))\Bigr]}
\biggr\}\right)
\lab{Z-qXX}
\end{equation}
We compare this formula with eq.(\ref{Z-R}) applied to the case of
cyclic $q$-XX-chain. More precise, we compare the perturbation series in
the hopping parameter $A$.

There are several technical simplifications to note. First of all, the
exponents $\E^{\eps_k(t_2-t_1)}$ may be excluded from (\ref{q-chron},
\ref{d}) since a generic term $\prod_{k=1}^n V(t_k)$ contains an equal
number of $\ad(t_k)$ and $a(t_k)$ and the energy on all sites are equal
to $B$. This leads to the cancelation of the exponents. Second point is
the fact that for this particular case the deformation parameter $q$
appears in formulae starting with  the order $M$ with respect to the
interaction (hopping). Indeed, for lower orders the only terms contribute
to the answer are the monomial terms with equal numbers of products
$\ad_ka_l$ and $\ad_la_k$ ($k=l\pm1$). When we collect such pairs of
products together using the permutation relations no phase factor
appears. This is because of the property of this products to commute with
terms as $\ad_m a_m$ ($m\ne k,l$).  When derivatives from expression
(\ref{s-fun}) act on $\prod\ad_ka_l\ad_la_k$ the deformation parameter
does not appear.  The result is natural: $m$-th order corresponds to $m$
hoppings. When $m<M$ there is no cyclic path for a particle to go and the
deformed statistics does not play a role~\cite{IK}.

Summing up, in the lower $M-1$ orders the contributions are pure
fermionic. It is interesting to compare nontrivial orders.  For any
particular finite $M$ it is possible to do by a straightforward
calculation. Here, for sake of simplicity, we present the simplest
nontrivial case $M=3$ since the case illustrate the situation by a very
clear way. First nontrivial order for the example is third order in the
hopping constant and it might be calculated immediately from
eq.(\ref{Z-R}):
\be
\frac{Z}{Z_0}=1+ 3\beta^2A^2[\bar{n}-{\bar{n}}^2]+
\beta^3A^3[-\bar{n}+{\bar{n}}^2(2+(q+q^{-1})/2)-
{\bar{n}}^3(1+(q+q^{-1})/2)] \ .
\lab{3por}
\ee
$\bar n$ is defined by eq.(\ref{bar-n}) and in this case is equal to
$\E^{-\beta B}/(1+\E^{-\beta B})$.

It is not difficult to see that the exact formula (\ref{Z-qXX})
gives the same result. Let us note that the expressions
in the round parentheses [$(2+(q+q^{-1})/2)$ and $(1+(q+q^{-1})/2)$] are
$q$-symmetrical coefficients and play the same role here as the standard
symmetrical coefficients do for the case $q=1$.
Technically, the parameter $q$ appears here from the permutation of
ordered vertices.

\section{Conclusion}
In the paper we derived the representation (in $q$-functional
derivatives) for the partition function and the Green's functions
generating functional starting with the Wick's theorems for creation and
annihilation operators of $q$-particles.  Derived representations may
be rewritten making use of functional integrals over functions on
($q$-deformed) graded-commutative space. It is not difficult to see that
this would bring us back to formulae of Ref.~\cite{IS}. Diagram technique
of the paper follows from both formalisms (functional derivatives and
functional integrals) equally easy. All said above let us state that the
quantum field theory machinery for particles with deformed exchange
statistics is constructed.  It is similar to the standard bosonic and
fermionic theories and differs by the $q$-deformed nature of fields ({\it
i.e.} corresponding classical variables) only.  We checked the technique
on an one-dimensional example.  The next step (which is much more
physically intriguing) is to examine a multi-dimensional case.
Long-range order, instabilities, influence of disorder might be some of
questions of interest for a such investigation.  We return to them in a
forthcoming paper.

\section*{Acknowledgments.}
We want to thank V.M. Agranovich for drawing of our attention to the
problems of quantum optics where $q$-particles find deep applications.
We are also grateful to A.N.Vasiliev for interesting discussions.
This work was supported by Grant of the Russian Fund of Fundamental
Investigations N 95-001-00548 and UK EPSRC grants GR/L29156 and GR/K68356.

\end{document}